\documentclass[%
 reprint,
 superscriptaddress,
 amsmath,amssymb,
 aps,
 prl,
]{revtex4-2}
\usepackage{graphicx}
\usepackage{dcolumn}
\usepackage{bm}
\usepackage{siunitx}
\usepackage{xcolor}
\usepackage{comment}
\usepackage{adjustbox}
\usepackage[utf8x]{inputenc}
\usepackage{ucs}
\usepackage[none]{hyphenat}
\usepackage{multirow}
\usepackage[normalem]{ulem}
\usepackage{subfigure}

\begin{document}
\title{
First Indication of Solar ${^8}$B Neutrino Flux through Coherent Elastic Neutrino-Nucleus Scattering in PandaX-4T
} 

\def\shKeyLab{School of Physics and Astronomy, Shanghai Jiao Tong University, Key Laboratory for Particle Astrophysics and Cosmology (MoE), Shanghai Key Laboratory for Particle Physics and Cosmology, Shanghai 200240, China}
\def\scKeyLab{Jinping Deep Underground Frontier Science and Dark Matter Key Laboratory of Sichuan Province}
\def\BUAA{School of Physics, Beihang University, Beijing 102206, China}
\def\BUAACenter{Peng Huanwu Collaborative Center for Research and Education, Beihang University, Beijing 100191, China}
\def\BUAALab{Beijing Key Laboratory of Advanced Nuclear Materials and Physics, Beihang University, Beijing 102206, China}
\def\SCNT{Southern Center for Nuclear-Science Theory (SCNT), Institute of Modern Physics, Chinese Academy of Sciences, Huizhou 516000, China}
\def\USTClab{State Key Laboratory of Particle Detection and Electronics, University of Science and Technology of China, Hefei 230026, China}
\def\USTCdep{Department of Modern Physics, University of Science and Technology of China, Hefei 230026, China}
\def\BUAALab{International Research Center for Nuclei and Particles in the Cosmos \& Beijing Key Laboratory of Advanced Nuclear Materials and Physics, Beihang University, Beijing 100191, China}
\def\pku{School of Physics, Peking University, Beijing 100871, China}
\def\YaLongSD{Yalong River Hydropower Development Company, Ltd., 288 Shuanglin Road, Chengdu 610051, China}
\def\IAP{Shanghai Institute of Applied Physics, Chinese Academy of Sciences, 201800 Shanghai, China}
\def\CHEPpku{Center for High Energy Physics, Peking University, Beijing 100871, China}
\def\SDUdep{Research Center for Particle Science and Technology, Institute of Frontier and Interdisciplinary Science, Shandong University, Qingdao 266237, China}
\def\SDUlab{Key Laboratory of Particle Physics and Particle Irradiation of Ministry of Education, Shandong University, Qingdao 266237, China}
\def\UMD{Department of Physics, University of Maryland, College Park, Maryland 20742, USA}
\def\TDLee{New Cornerstone Science Laboratory, Tsung-Dao Lee Institute, Shanghai Jiao Tong University, Shanghai 201210, China}
\def\MESJTU{School of Mechanical Engineering, Shanghai Jiao Tong University, Shanghai 200240, China}
\def\SYU{School of Physics, Sun Yat-Sen University, Guangzhou 510275, China}
\def\SYUSFI{Sino-French Institute of Nuclear Engineering and Technology, Sun Yat-Sen University, Zhuhai 519082, China}
\def\NKU{School of Physics, Nankai University, Tianjin 300071, China}
\def\YTU{Department of Physics, Yantai University, Yantai 264005, China}
\def\FDU{Key Laboratory of Nuclear Physics and Ion-beam Application (MOE), Institute of Modern Physics, Fudan University, Shanghai 200433, China}
\def\USST{School of Medical Instrument and Food Engineering, University of Shanghai for Science and Technology, Shanghai 200093, China}
\def\SJTUSC{Shanghai Jiao Tong University Sichuan Research Institute, Chengdu 610213, China}
\def\SPEIT{SJTU Paris Elite Institute of Technology, Shanghai Jiao Tong University, Shanghai 200240, China}
\def\NNU{School of Physics and Technology, Nanjing Normal University, Nanjing 210023, China}
\def\SYSUzhuhai{School of Physics and Astronomy, Sun Yat-Sen University, Zhuhai 519082, China}
\def\CDUT{College of Nuclear Technology and Automation Engineering, Chengdu University of Technology, Chengdu 610059, China}

\affiliation{\TDLee}
\author{Zihao Bo}\affiliation{\shKeyLab}
\author{Wei Chen}\affiliation{\shKeyLab}
\author{Xun Chen}\affiliation{\TDLee}\affiliation{\shKeyLab}\affiliation{\SJTUSC}\affiliation{\scKeyLab}
\author{Yunhua Chen}\affiliation{\YaLongSD}\affiliation{\scKeyLab}
\author{Zhaokan Cheng}\affiliation{\SYUSFI}
\author{Xiangyi Cui}\affiliation{\TDLee}
\author{Yingjie Fan}\affiliation{\YTU}
\author{Deqing Fang}\affiliation{\FDU}
\author{Zhixing Gao}\affiliation{\shKeyLab}
\author{Lisheng Geng}\affiliation{\BUAA}\affiliation{\BUAACenter}\affiliation{\BUAALab}\affiliation{\SCNT}
% ...
\collaboration{PandaX Collaboration}
\noaffiliation

\date{\today}

\begin{abstract}
The PandaX-4T liquid xenon detector at the China Jinping Underground Laboratory is used to measure the solar $^8$B neutrino flux by detecting neutrinos through coherent scattering with xenon nuclei.
Data samples requiring the coincidence of scintillation and ionization signals (paired), as well as unpaired ionization-only signals (US2), are selected with energy threshold of approximately 1.1~keV (0.33~keV) nuclear recoil energy. 
Combining the commissioning run and the first science run of PandaX-4T, a total exposure of 1.20 and 1.04~tonne$\cdot$year are collected for the paired and US2, respectively. 
After unblinding, 3 and 332 events are observed with an expectation of 2.8$\pm$0.5 and 251$\pm$32 background events, for the paired and US2 data, respectively. 
A combined analysis yields a best-fit $^8$B neutrino signal of 3.5 (75) events from the paired (US2) data sample, with $\sim$37\% uncertainty, and the  background-only hypothesis is disfavored at 2.64~$\sigma$ significance.
This gives a solar $^8$B neutrino flux of ($8.4\pm3.1$)$\times$10$^6$\,cm$^{-2}$s$^{-1}$, consistent with the standard solar model prediction. 
It is also the first indication of solar $^8$B neutrino ``fog'' in a dark matter direct detection experiment. 
\end{abstract}

\maketitle

\newcommand{\mwba}[1]{\textcolor{violet}{#1}}
\newcommand{\mwbd}[1]{\textcolor{violet}{\sout{#1}}}

In 1984, Drukier and Stodolsky proposed to detect low-energy neutrinos via coherent elastic neutrino-nucleus scattering (CE$\nu$NS)~\cite{PhysRevD.30.2295}, 
a neutral current interaction first pointed out by Freedman~\cite{freedman1974coherent}.
With a typical momentum transfer of MeV/$c^2$, the neutral current interaction with individual nucleons is coherently enhanced. The main experimental challenge, however, is the detection of ultra-low-energy nuclear recoil events at the keV-scale.
Goodman and Witten soon realized that such low energy neutral current detectors are ideal for searching for the scattering of dark matter (DM) particles with nuclei~\cite{goodman1985detectability}, i.e. the DM direct detection. In 2017, CE$\nu$NS was first observed with accelerator neutrinos scattering on a CsI detector~\cite{akimov2017observation}, followed by several more precise measurements~\cite{akimov2019first, akimov2021first, adamski2024first}. On the other hand, despite tremendous progress in the sensitivity of DM direct detection experiments over the years, no convincing DM signals have emerged. 
It is widely anticipated that these DM detectors will eventually observe the CE$\nu$NS induced by neutrinos of both astronomical and terrestrial origins, for example, the solar neutrinos and atmospheric neutrinos. These irreducible background neutrinos, now referred to as the ``neutrino fog'', will ultimately impose limitations on the sensitivity of DM detection~\cite{billard2014implication, o2021new}, but they hold significant scientific value in their own right. Current generation multi-tonne liquid xenon experiments~\cite{meng2021dark, aprile2023first, aalbers2023first} are racing to observe solar $^8$B CE$\nu$NS signals. 
A detection would not only be the first of its kind, 
but also complementary to those earlier measurements of solar $^8$B neutrinos by large neutrino detectors such as SNO and SNO+~\cite{SNO:2011hxd, anderson2019measurement}, Super Kamiokande~\cite{Super-Kamiokande:2016yck}, Borexino~\cite{agostini2020improved}, and KamLAND~\cite{abe2011measurement}.
It would further empower these DM detectors to become unique astrophysical neutrino observatories, particularly in detecting supernova neutrino bursts~\cite{burrows1992future,  lang2016supernova, aalbers2022next, pang2024detecting}. 
It can also provide a wealth of particle physics opportunities such as the measurement of weak mixing angle at low momentum transfer~\cite{canas2018future}, and searching for possible non-standard neutrino interactions~\cite{barranco2007sensitivity} or sterile neutrinos~\cite{Anderson:2012pn}. Furthermore, CE$\nu$NS detection technology may find important applications in nuclear safeguards~\cite{von2022use}.

The PandaX-4T experiment, located at the China Jinping Underground Laboratory~\cite{kang2010status, li2015second}, utilizes a dual-phase 
time-projection-chamber (TPC) filled with 3.7~tonne of liquid xenon, with an initial primary goal to detect DM particles.
The TPC is equipped with 169 and 199 3-inch Hamamatsu R11410 photomultiplier tubes (PMTs) on the top and bottom, respectively, detecting the scintillation signal ($S1$) and ionization signal ($S2$).
Detailed descriptions of PandaX-4T can be found in Refs.~\cite{he2021500, yan2021pandax, zhao2021cryogenics, meng2021dark}. 

Both XENON1T and PandaX-4T have previously reported limits of the solar $^8$B neutrino flux~\cite{aprile2021search, ma2023search}. 
In this work, we report an updated search utilizing both the commissioning run (Run0) and the first science run (Run1) from PandaX-4T, with a data-taking time of 95 and 164 calendar days, respectively.
During the two periods, the TPC was operated with an electric field of approximately 93 and 84\,V/cm, respectively.
The scintillation photon detection efficiency ($g1$) is approximately 10\%, and the product of the extraction efficiency and amplification is between 17 and 20 photoelectrons (PEs) per ionized electron ($g2$)~\cite{luo2024signal}.



Different from the previous publications~\cite{meng2021dark, ma2023search, li2023search, pandax2023limits, huang2023search}, we reblinded Run0 data and carried out a blind analysis of both data sets using an improved event building procedure, as detailed in Ref.~\cite{luo2024signal}. 
The data selection combines two distinct event classes, the paired $S1$ and $S2$ signals (paired), as well as isolated unpaired $S2$ signals (US2).
The paired data benefit from effective electronic recoil (ER) and nuclear recoil (NR) discrimination and three-dimensional position reconstruction capability, but at the cost of a higher energy threshold.
Conversely, the US2 data allow for a lower energy threshold, but with the price of a higher background rate due to the absence of knowledge of the vertical position.

The basic data set used in this analysis is the same as in the DM search~\cite{run1wimpsearch} by excluding approximately 2.3 days (Run0) and 2.2 days (Run1) with unstable operation conditions. 
In addition, data periods of 5.5 days (paired) and 29.1 days (US2) are excluded due to  
abnormal $S1$ and $S2$ rates. 
As reported in previous work~\cite{meng2021dark,ma2023search}, a veto cut is applied to the time window after any signals exceeding 10,000\,PE (see Ref.~\cite{luo2024signal} for more details). For US2, an additional volume cut within an 80~mm radius cylinder surrounding the previous large $S2$ ($>$20,000\,PE) is also imposed.
These measures are implemented to mitigate the ``afterglow'' effect, also known as delayed electrons~\cite{akerib2020investigation, sorensen2017two, sorensen2017electron, akerib2021improving} in the literature.
The afterglow veto is optimized independently for the paired data and US2 data, resulting in an exposure loss of 23\% (34\%) and 26\% (33\%) in Run0 (Run1), respectively. 

\begin{figure}[htp]
    \centering
    \includegraphics[width=0.95\columnwidth]{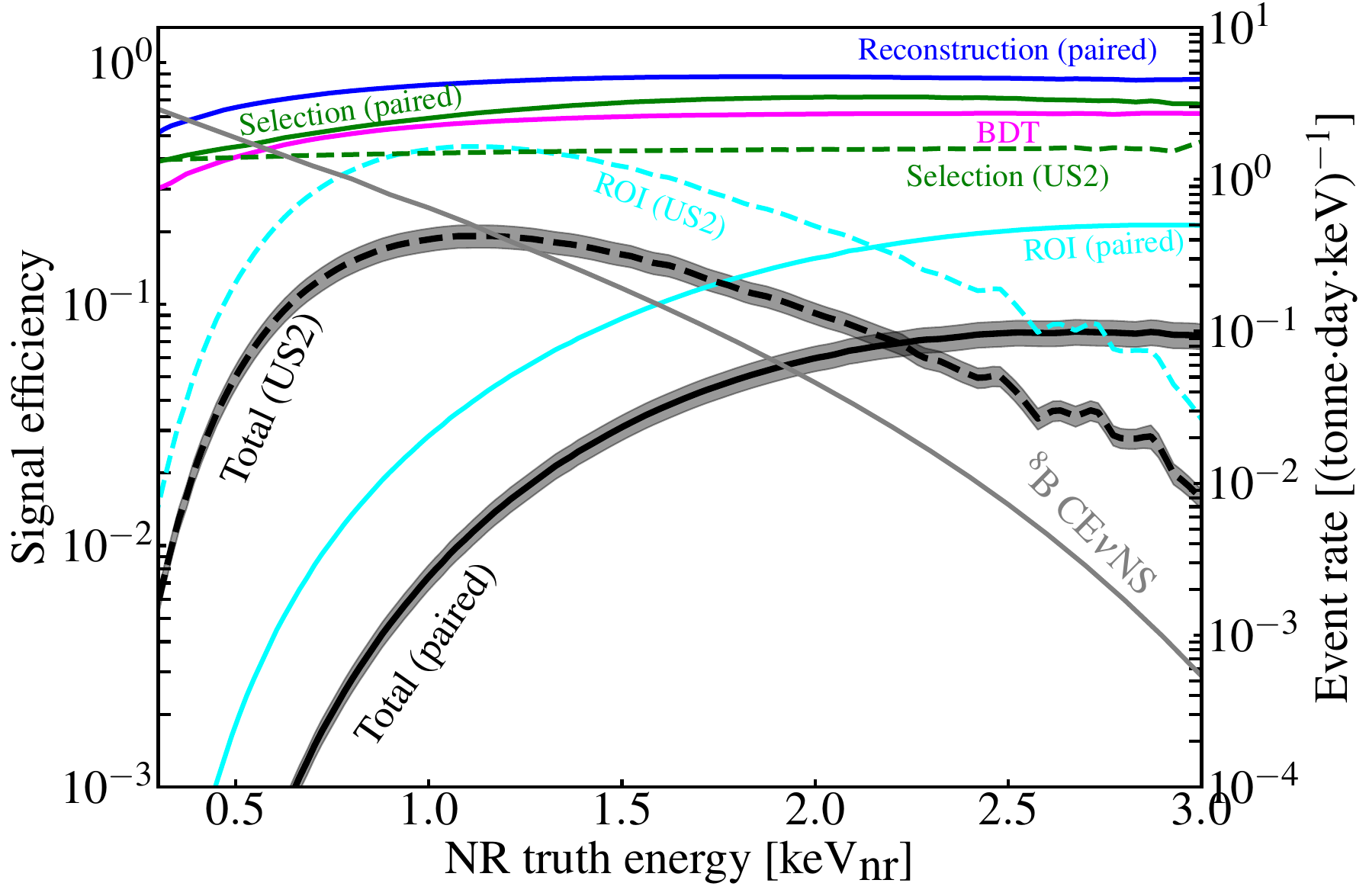}
    \caption{
    All components of $^8$B candidate selection efficiency for the paired (solid) and US2 (dashed) data: green=data selection, blue=signal reconstruction, cyan=ROI, magenta=BDT, black=total.
    The gray shaded regions indicate $\pm$1$\sigma$ uncertainties.
    $^8$B CE$\nu$NS signal spectrum is overlaid in gray with the scale indicated on the right axis.
    If using 1\% efficiency as a threshold, the corresponding NR energy threshold for the paired and US2 are 1.1 and 0.33\,keV, respectively.
    }
    \label{fig:efficiency}
\end{figure}

The $^8$B candidate selection has three main steps: the data selection, the signal reconstruction, and the region-of-interest (ROI) selection.
The data selection cuts for the paired and US2 data mostly follow those in Refs.~\cite{ma2023search,li2023search}.
For US2, the data selection cuts are further loosened to gain higher acceptance.
The signal reconstruction refers to finding the correct paired and US2 data.
For the paired data, inefficiency exists since some $S1$ and $S2$ could be mis-paired due to the existing spurious $S1$s in the event window~\cite{ma2023search}. 
For the US2 data, the signal reconstruction follows that in Ref.~\cite{li2023search} with negligible efficiency loss. 
The ROI for the paired data is defined by requiring $S1$ signals with two or three ``fired'' PMT hits and the raw $S2$ signals in the range between 60 to 300\,PE.
The ROI of the US2 data requires no isolated $S1$ signal with 2 or more hits in the event window, and the $S2$ corresponds to 4 to 8 ionized electrons.
The $S2$ range is different from Ref.~\cite{li2023search}, to minimize the influence of the micro-discharging (MD) background in the low-energy region (see later).
The efficiencies of these three components, shown in Fig.~\ref{fig:efficiency}, are estimated using dedicated waveform simulation (WS) algorithm~\cite{lin2024waveform}.

\begin{figure}
    \centering
    \includegraphics[width=1\linewidth]{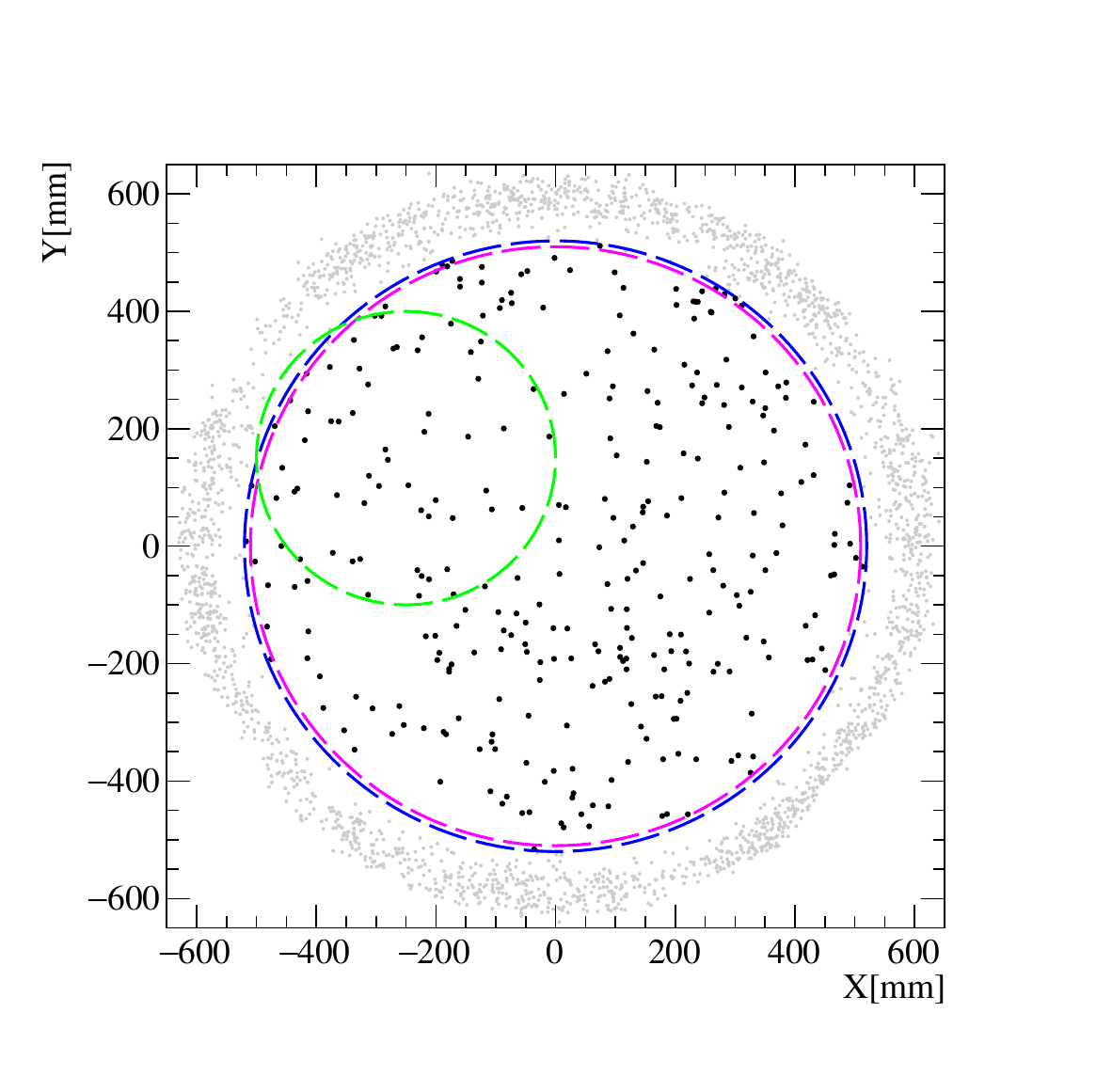}
    \caption{Unblinded US2 data within (black) and outside (gray) the FV. The magenta and blue circles correspond to Run0 and Run1 FV circle cuts, respectively, and the green circle represents an additional cylindrical cut in Run1 due to dysfunctional PMTs.}
    \label{fig:US2_FV}
\end{figure}

The fiducial volume (FV) selections for Run0 and Run1 are separately optimized based on the background models in each run.
For paired data, a radius cut of 520\,mm is applied to both runs. 
For US2 data, we set a radius cut of 510\,mm for Run0 and 520\,mm for Run1. We also remove a cylindrical region within a radius of 250\,mm in the upper-left corner  (Fig.~\ref{fig:US2_FV})  in Run1 due to a number of dysfunctional top PMTs~\cite{run1wimpsearch}.
The fiducial masses of Run0 and Run1 are 2.58 (2.78) and 2.55 (2.16)\,tonne, respectively, for the paired (US2) data.
In combination with the live time from each run, including the cylindrical after-glow veto for US2, the total exposure is 1.20 (paired) and 1.04 (US2)~tonne$\cdot$year.

\begin{table*}[htp]
    \centering
    \begin{tabular}{cc|cc|cc}
    \hline\hline
        &
        &
    \multicolumn{2}{c|}{Run0}    &
    \multicolumn{2}{c}{Run1}    \\
        &
        &
    W/O BDT  &
    W/ BDT   &
    W/O BDT  &
    W/ BDT   \\
    \hline\hline
    \multirow{2}{*}{Off-window}      &
    Prediction          &
    180$\pm$27      &
    0.9$\pm$0.2     &
    417$\pm$63     &
    1.2$\pm$0.4     \\
          &
    Data   &
    205     &
    1     &
    404     &
    0     \\ \hline
    \multirow{2}{*}{10\% data}          &
    Prediction  &
    26$\pm$6   &
    0.12$\pm$0.04     &
    34$\pm$7     &
    0.06$\pm$0.02     \\
              &
    Data  &
    18     &
    0     &
    29     &
    0     \\ \hline
    \multirow{2}{*}{1-hit side-band}      &
    Prediction   &
    17095$\pm$2564     &
    14$\pm$4     &
    27567$\pm$4135     &
    15$\pm$5   \\
          &
    Data   &
    17374   &
    9     &
    29359    & 
    17     \\
    \hline\hline
    \end{tabular}
    \caption{
    Summary of predicted and observed numbers of events in the off-window, $\sim$10\% unblinded data, and 1-hit sideband data, before and after the BDT.
    }
    \label{tab:ac_validation}
\end{table*}

\begin{figure}[htp]
    \centering
     \includegraphics[width=0.9\columnwidth]{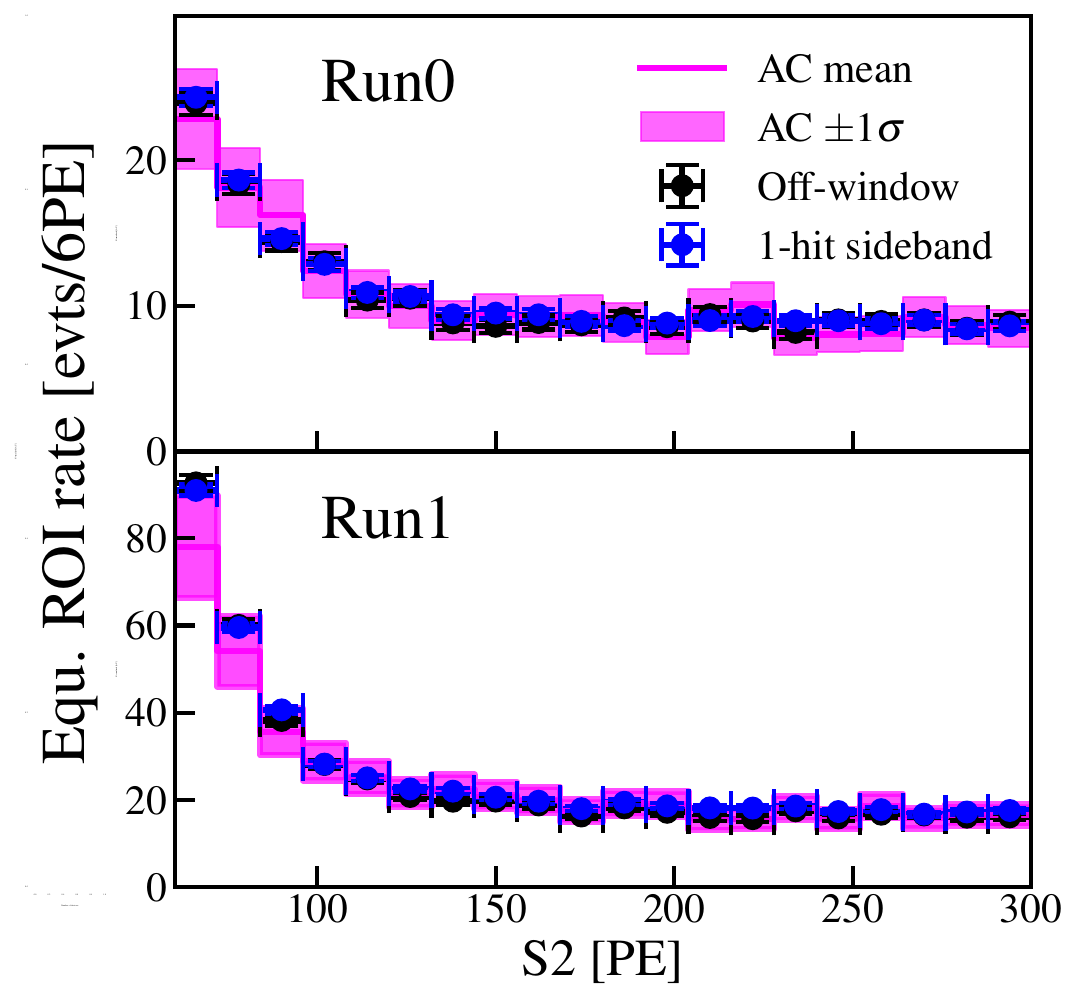}
    \caption{
    Comparison of the normalized $S2$ spectra between the AC model (magenta histogram with shaded uncertainty), off-window data (black), and the 1-hit sideband data (blue).
    }
    \label{fig:validation_compare}
\end{figure}

The dominant background in the paired data is the accidental coincidence (AC) background, which arises from random pileup of isolated $S1$ and $S2$ signals. 
Other types of background, such as neutron-induced NRs, $\gamma$/$\beta$-induced ERs, and surface backgrounds, are negligible.
To model the AC background, the selected isolated $S1$ and $S2$ waveforms from the real data are randomly scrambled and ``stitched'' together.
These scrambled waveforms are then processed through the same reconstruction algorithms as in the real data. 
Such an AC model is validated using three types of data in Table~\ref{tab:ac_validation}: the events with drift time larger than the maximum value allowed by the TPC (off-window data), paired data selected from a randomly-chosen 10\% calendar time (``10\% data''), and all paired data with $S1$ consisting of a single PMT hit (1-hit sideband).
The $S2$ spectra of the AC model prediction and the off-window and 1-hit sideband validation data are compared in Fig.~\ref{fig:validation_compare}, based on which an overall 15\% uncertainty is assigned. 
 A Boosted Decision Tree (BDT) algorithm is developed to further reject AC background~\cite{ma2023search}, in which the signal training samples are $^8$B events simulated by the signal response model~\cite{luo2024signal} and the WS~\cite{lin2024waveform}.  
The BDT cuts are optimized separately for each run to have an optimal signal-to-background ratio in the ROI.
Its rejection power against AC mainly comes from the $S2$ related variables.
The predicted AC rates for these validation datasets, both before and after the BDT cut, are summarized in Table~\ref{tab:ac_roi_prediction}, where good agreements are found. 

\begin{table}[htp]
    \centering
    \begin{tabular}{c|cc|cc}
    \hline\hline
        &
    \multicolumn{2}{c|}{Run0}    &
    \multicolumn{2}{c}{Run1}    \\
    paired ROI    &
    2-hit           &
    3-hit           &
    2-hit           &
    3-hit           \\
    \hline\hline
    Surface           &
    0.06$\pm$0.01   &
    0.06$\pm$0.01   &
    0.01$\pm$0.01   &
    0.02$\pm$0.02   \\
    ER              &
    0.01$\pm$0.00   &
    0.00$\pm$0.00   &
    0.01$\pm$0.01   &
    0.01$\pm$0.01   \\
    Neutron         &
    0.02$\pm$0.01   &
    0.02$\pm$0.01   &
    0.03$\pm$0.01   &
    0.03$\pm$0.01   \\
    AC              &
    1.08$\pm$0.28   &
    0.07$\pm$0.02   &
    1.15$\pm$0.35   &
    0.24$\pm$0.08   \\ \hline
    Total bkg.    & 
    1.16$\pm$0.28       &
    0.15$\pm$0.02       &
    1.21$\pm$0.35       &
    0.30$\pm$0.08       \\
    $^8$B CE$\nu$NS     &
    1.00$\pm$0.24       &
    0.24$\pm$0.09       &
    1.76$\pm$0.50       &
    0.40$\pm$0.18       \\
    \hline
    Observed            &
    1                   &
    0                   &
    2                   &
    0                   \\
    \hline\hline
    \end{tabular}
    \caption{
    Paired data: Run0 and Run1 expected events in 2- and 3-hit regions from the surface, ER, neutron, and the AC background, and the $^8$B CE$\nu$NS signal. The observed events after unblinding are given in the last row.
    }
    \label{tab:ac_roi_prediction}
\end{table}

In the ROI of the US2 data, the radioactivity on the cathode electrode (cathode background) contributes to $\sim$80\% of the background.
Similar to the approach in Ref.~\cite{li2023search}, the cathode background spectral shape is estimated using the data sample with both $S1$ ($<$50 PE) and $S2$ signals identified, and a characteristic drift time matching the maximal drift time of the TPC (cathode control sample). 
No significant $S1$ dependence on the $S2$ spectral shape is observed. Therefore, for US2, we assume the cathode background has the same shape as the cathode control sample.
The cathode background rate is estimated using US2 data but in a sideband range between 11 to 15 ionized electrons.
The ``micro-discharging'' (MD) background is the second-largest background in the ROI, with a sharply-decreasing spectral shape at the low end. 
The MD background shows a clear time-dependence, and has a particularly high rate in set3 of Run0, which was used as the spectral template in the previous work~\cite{li2023search}. 
In this analysis, we observe that the events rejected by the afterglow veto have consistent distributions in $S2$ charge and $S2$ width compared to the previous MD model. Due to the larger statistics, we use these afterglow vetoed events to estimate the shape of the MD.
The MD background rate is estimated using the US2 data but within an $S2$ sideband range between 2.5 to 4 ionized electrons.
Within the ROI, the ER background is estimated to contribute less than 1\% of the total background.

\begin{table}[htp]
    \centering
    \begin{tabular}{c|cc}
    \hline\hline
    US2 ROI    &
    Run0    &
    Run1    \\
    \hline\hline
    Cathode &
    100$\pm$24&
    104$\pm$21\\
    MD      &
    25$\pm$3&
    20$\pm$4\\ 
    ERs     &
    1.3$\pm$0.1&
    0.9$\pm$0.2\\
    \hline
    Total bkg.  &
    126$\pm$24&
    125$\pm$21\\
    $^8$B CE$\nu$NS  &
    18$\pm$4&
    25$\pm$6 \\
    \hline
    Observed    &
    158&
    174\\
    \hline\hline
    \end{tabular}
    \caption{
    US2 data: Run0 and Run1 expected events from the cathode, MD, and ER background and $^8$B CE$\nu$NS signal in the ROI.  The observed events after unblinding are given in the last row.
    }
    \label{tab:us2_background}
\end{table}


 
\begin{table*}[htp]
    \centering 
    \begin{tabular}{cccccccc}
    \hline\hline
        &
    \multicolumn{4}{c}{Paired}  &
    \multicolumn{2}{c}{US2}      &
        \\
        &
      \multicolumn{2}{c}{Run0}  &
      \multicolumn{2}{c}{Run1}  &
      Run0  &
      Run1  &
        \\
    Sources &
    2-hit &
    3-hit   &
    2-hit  &
    3-hit   &
      &
         &
    Estimated by    \\
    \hline\hline
    $^8$B data selection           &
    \multicolumn{2}{c}{0.10}                 &
    \multicolumn{2}{c}{0.10}                &
    0.11                 &
    0.17                 &
    WS vs. DS           \\
    $^8$B model&
    0.24                 &
    0.37                 &
    0.28                 &
    0.44                 &
    0.17                 &
    0.16                 &
    Prescribed NEST model uncertainty        \\
    BDT to $^8$B     &
    \multicolumn{2}{c}{0.17}                 &
    \multicolumn{2}{c}{0.11}                 &
    -                   &
    -                   &
    WS vs. DS           \\
    \hline
    AC model            &
    \multicolumn{2}{c}{0.15}                &
    \multicolumn{2}{c}{0.15}                 &
    -                   &
    -                   &
    scrambled AC samples vs. control samples  \\
    BDT to AC           &
    \multicolumn{2}{c}{0.18}                 &
    \multicolumn{2}{c}{0.25}                 &
    -                   &
    -                   &
    scrambled AC samples vs. control samples  \\
    Cathode model       &
    -                   &
    -                   &
    -                   &
    -                   &
    0.24                 &
    0.20                &
    Varying side-band selection \\
    MD model            &
    -                   &
    -                   &
    -                   &
    -                   &
    0.13                 &
    0.16                 &
    Varying side-band selection \\
    \hline\hline
    \end{tabular}
    \caption{
    Summary of systematic uncertainties included in the combined likelihood fit, with first three rows on the $^8$B signals and next four on backgrounds. 
    The methods to estimate uncertainties are given in the last column. See text for details.
    All values are dimensionless, defined as the induced fractional change in the signal or background rate in the ROI.
    }
    \label{tab:systematics}
\end{table*}

\begin{figure}
    \centering
    \includegraphics[width=0.9\columnwidth]{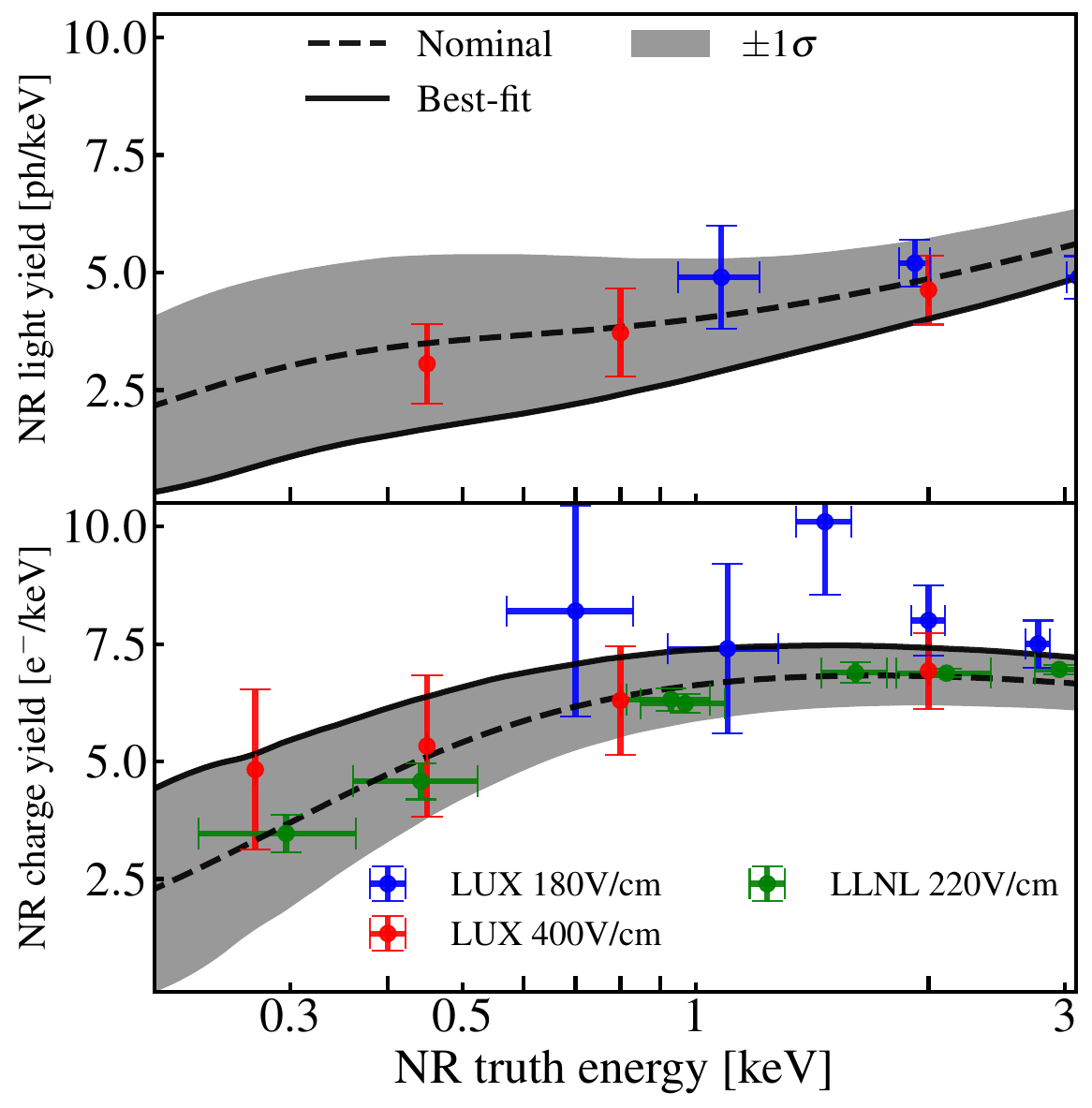}
    \caption{
    The light (top) and charge (bottom) yields as a function of NR truth energy from NEST-2.3.6~\cite{szydagis2018noble}, with median (dashed curves) and $\pm$1$\sigma$ bands (shaded)~\cite{szydagis2022review}.
    The solid lines represent the best-fits after the combined likelihood fit using our data.
    The yields for Run0 and Run1 have negligible differences.
    The measurements (colored data points) from~\cite{akerib2016low, huang2020ultra, lenardo2019measurement} taken at different fields are overlaid for reference.
    }
    \label{fig:lycy}
\end{figure}

The systematic uncertainties of the $^8$B CE$\nu$NS signals in both paired and US2 data are summarized in Tables~\ref{tab:systematics}. There are three major contributions: the data selection efficiency, the $^8$B signal model in liquid xenon, and the BDT cut efficiency (paired only). 
The uncertainty of the data selection and BDT efficiencies to $^8$B signal are estimated by comparing the efficiencies evaluated using the WS and those obtained from the double-scatter (DS) neutron calibration data. Only the secondary $S2$s in the DS samples are selected to gain statistics in the ROI.
For the $^8$B signal model, we begin with the differential flux of the solar $^8$B neutrinos given in Ref.~\cite{bahcall2005new, bahcall1996standard}.
It is worth mentioning that a recent update~\cite{acharya2024solar} indicates an approximately 5\% difference in the spectral shape, non-significant in this study, but may become important as the experimental precision improves.
The deposit NR energy spectrum in liquid xenon is computed by folding the neutrino flux with the standard model CE$\nu$NS cross section in Ref.~\cite{ruppin2014complementarity}, then converted to the expected $S1$ and $S2$ distributions in the paired and US2 data using the NESTv2.3.6-based~\cite{szydagis2018noble} signal model~\cite{luo2024signal}.
NEST is a global fit of all available direct measurements of light and charge production vs. NR energy in liquid xenon. 
Several components of fluctuations are parameterized and incorporated in the model, dominated by the Poisson and binomial fluctuations in the quanta production and detection within the ROI.
The uncertainties in the NEST model~\cite{szydagis2022review} (Fig.~\ref{fig:lycy}), thereby the induced variations in the means and fluctuations of the $S1$ and $S2$, are automatically carried through by our signal model~\cite{luo2024signal} to predict the change of $^8$B events~\footnote{Using the latest version of NEST (v2.4.0) makes negligible difference in both the signal spectral shape and rate compared to NEST v2.3.6 we used.}.
The translated systematic uncertainties to $^8$B neutrino rate are approximately 24\% (Run0) and 28\% (Run1) for 2-hit, and 37\% (Run0) and 44\% (Run1) for the 3-hit data, correlated between the 2-hit and 3-hit. The corresponding uncertainty for US2 is 17\%, anti-correlated with the paired data.


\begin{figure}[htp]
    \centering
    \includegraphics[width=0.99\columnwidth]{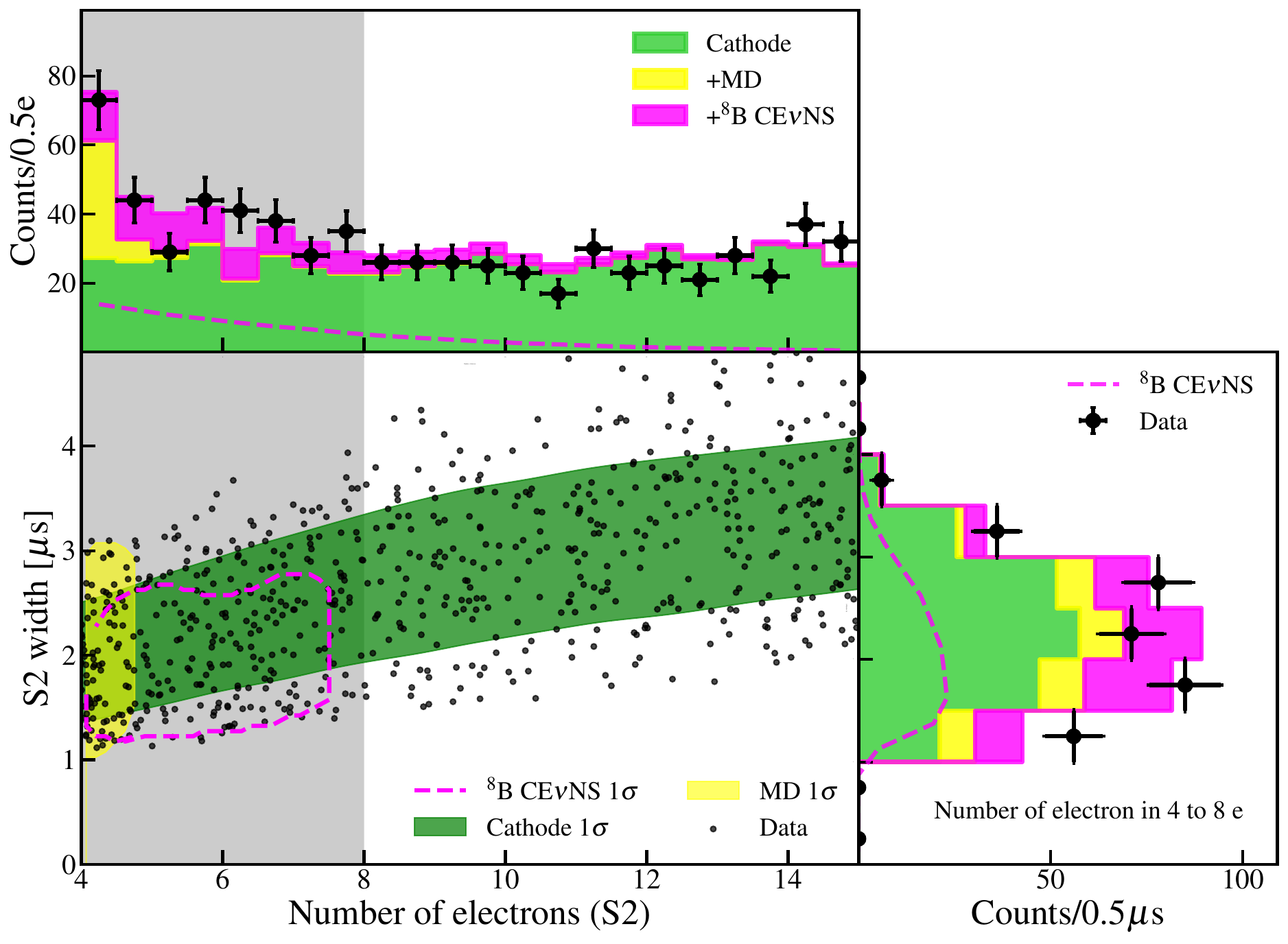}
    \caption{
    Unblinded US2 in Run0 and Run1 plotted in $S2$ (converted to number of ionized electrons) vs. $S2$ width, overlaid with 1$\sigma$ contours of the cathode (green), MD (yellow), and $^8$B in the ROI (magenta dashed). Gray shaded region indicates the ROI. 
    The top and right panels show one-dimensional projections in $S2$ and $S2$-width, with stacked best-fit contributions from the cathode (green), MD (yellow), and $^8$B (magenta). 
    The dashed magenta curves (unstacked) represent the best-fit $^8$B distributions. 
    }
    \label{fig:us2_data}
\end{figure}

The systematic uncertainties of the background are also summarized in Table~\ref{tab:systematics}.
For the background in the paired data, the AC model uncertainty (pre-BDT) has been estimated to be 15\% (Fig.~\ref{fig:validation_compare}). 
The BDT efficiency uncertainty for the AC is determined based on the difference of BDT responses to the scrambled AC samples and 1-hit sideband data.
The systematic uncertainties of the cathode and MD background in the US2 data are both given as the root-mean-square of the predictions when varying the selection criteria for the sideband data sample.


The ROI of the paired data  and the US2 data (excluding the 10\% data) have been blinded until the data selections and the background models are settled.
After unblinding, 1 (0) and 2 (0) candidate events are observed in Run0 and Run1, respectively, for the 2-hit (3-hit) data samples (Table~\ref{tab:ac_roi_prediction}).
The single 2-hit candidate event in Run0 is confirmed to be the same as Ref.~\cite{ma2023search}.
In total, 332 events are observed in the US2 data after unblinding, and their $S2$ charge and width distribution is
shown in Fig~\ref{fig:us2_data}.


The statistical interpretation is performed based on a two-sided profile likelihood ratio (PLR) method~\cite{cowan2011asymptotic} with a combined binned likelihood between the paired and US2 data $\mathcal{L} = \mathcal{L}_{\textrm{paired}} \cdot \mathcal{L}_{\textrm{US2}}$, where
$\mathcal{L}_{\textrm{paired}}$ and $\mathcal{L}_{\textrm{US2}}$ are constructed similarly as in Refs.~\cite{ma2023search, li2023search}. 
The likelihood of US2 is constructed on a two-dimensional space of $S2$ charge and $S2$ width, with corresponding distributions illustrated in Fig.~\ref{fig:us2_data}.
The $^8$B $S2$s are distributed as a slow exponential spectrum, different from the flat spectrum of the cathode and the sharply decreasing spectrum of the MD.
The cathode events have a larger average width compared to the $^8$B events due to the diffusion effect after a long drift from the bottom of the TPC.
All systematic uncertainties in Table~\ref{tab:systematics} are treated as uncorrelated nuisance parameters, except those for the $^8$B signal model, which is implemented as global in both runs, but anti-correlated between the paired and US2 data. The $^8$B signal strength is the only floating fit parameter.

\begin{figure}[htp]
    \centering
    \includegraphics[width=0.95\columnwidth]{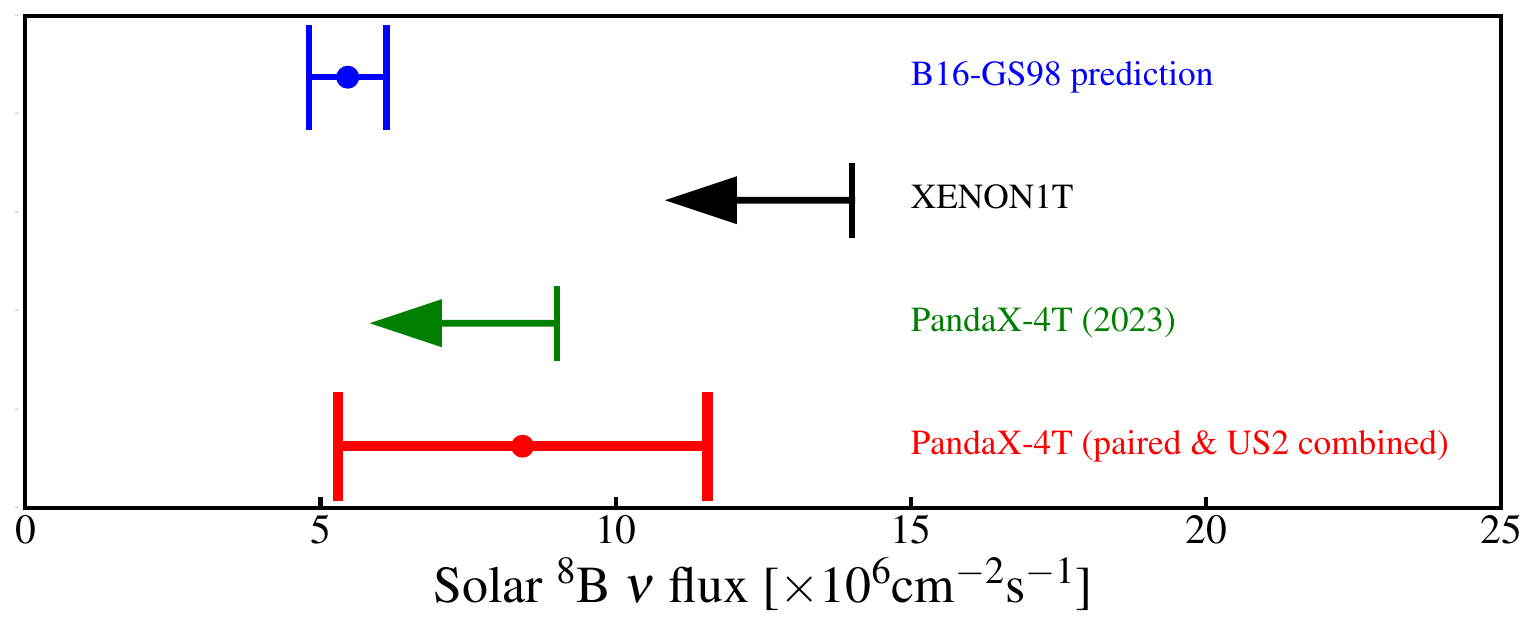}
    \caption{
    The best-fit $^8$B solar neutrino flux and 1$\sigma$ uncertainty from this work (red), together with 90\% C.L. regions of the PandaX-4T previous constraint~\cite{li2023search} (green), XENON1T constraint~\cite{aprile2021search} (black), and 1$\sigma$ of the theoretical prediction from the standard solar model~\cite{vinyoles2017new} (blue).
    }
    \label{fig:flux}
\end{figure}

The joint likelihood fit gives a best-fit $^8$B events of 75$\pm$28 (US2), and 3.5$\pm$1.3 (paired), with uncertainties fully correlated. The one-dimensional projections of the data and best-fit in $S2$ charge and width are shown in Fig.~\ref{fig:us2_data}, where good agreements are observed with a goodness-of-fit p-value of 0.31. 
For comparison, with US2 data alone, the best-fit $^8$B is 92$\pm$34 events. 
All best-fit nuisance parameters are within $\pm1\sigma$ from the nominal values in Table~\ref{tab:systematics}. 
The best-fit charge yield curve is at $+0.97\sigma$ (Fig.~\ref{fig:lycy}),
representing the largest pull among all systematic uncertainties.
The PLR analysis shows that our data have a p-value of 0.004 (2.64$\sigma$ significance) against the background-only hypothesis.
The resulting best-fit solar $^8$B neutrino flux, summarized in Fig.~\ref{fig:flux}, is (8.4$\pm$3.1)$\times$10$^6$\,cm$^{-2}$s$^{-1}$, consistent with the standard solar model prediction~\cite{vinyoles2017new}.


To summarize, we have performed an updated search of the solar $^8$B neutrino using the CE$\nu$NS events in the PandaX-4T experiment.
Both the paired and US2 data in Run0 and Run1 are utilized, leading to nuclear recoil energy thresholds of 1.1 and 0.33 keV, and a total exposure of 1.20 and 1.04~tonne$\cdot$year, respectively.
The best fit $^8$B events are 75$\pm$28 (US2), and 3.5$\pm$1.3 (paired), which disfavors background-only hypothesis at 2.64$\sigma$. 
The measured $^8$B neutrino flux is (8.4$\pm$3.1)$\times$10$^6$\,cm$^{-2}$s$^{-1}$, consistent with the prediction of the standard solar model~\cite{vinyoles2017new} within 1$\sigma$ uncertainty.
Our result represents the first positive indication of solar $^8$B neutrino fog in a dark matter direct detection experiment. With ongoing data taking of PandaX-4T and the future upgrades~\cite{PandaX:2024oxq}, we expect to significantly improve the $^8$B neutrino measurement, opening exciting new science opportunities from deep underground. 
The development of liquid xenon detectors may also open new avenues for low-energy neutrino detection using the CE$\nu$NS channel. 


\paragraph{Author contributions}
This work is the result of the contributions and efforts of all participating institutes of the PandaX Collaboration, under the leadership of the hosting institute, Shanghai Jiao Tong University. The collaboration has constructed and operated the PandaX-4T apparatus, and performed the data processing, calibration and data selections. J. Liu is the Collaboration Spokesperson. M. Zhang, J. Li, Z. Qian performed the data analysis and hypothesis test under the guidance of N. Zhou, Q. Lin and Y. Meng. M. Zhang  mainly analyzed the ionization-only data and background, J. Li mainly analyzed the paired data and background, Z. Qian mainly studied the signal yield and performed the statistical analyses. The paper draft was prepared by Q. Lin and N. Zhou, extensively edited by J. Liu. All authors approved the final version of the manuscript.


 

This project is supported in part by grants from National Science Foundation of China (Nos. 12090060, 12090061, 12090063, 12325505, 12222505, 12275267, U23B2070), a grant from the Ministry of Science and Technology of China (Nos. 2023YFA1606200, 2023YFA1606201), and by Office of Science and Technology, Shanghai Municipal Government (grant Nos. 22JC1410100, 21TQ1400218), Sichuan Province Innovative Talent Funding Project for Postdoctoral Fellows (No. BX202322), Sichuan Provincial Natural Science Foundation (No. 2024NSFSC1374). We thank for the support by the Fundamental Research Funds for the Central Universities. We also thank the sponsorship from the Chinese Academy of Sciences Center for Excellence in Particle Physics (CCEPP), Hongwen Foundation in Hong Kong, New Cornerstone Science Foundation, Tencent Foundation in China, and Yangyang Development Fund. Finally, we thank the CJPL administration and the Yalong River Hydropower Development Company Ltd. for indispensable logistical support and other help. 



Note added: as we are finalizing the revision of this manuscript after a round of peer review, we also notice the preliminary results of the $^8$B flux measurement from XENONnT collaboration with a similar significance in a preprint~\cite{aprile2024first}.

\begin{figure}[htp]
    \centering
    \includegraphics[width=0.95\columnwidth]{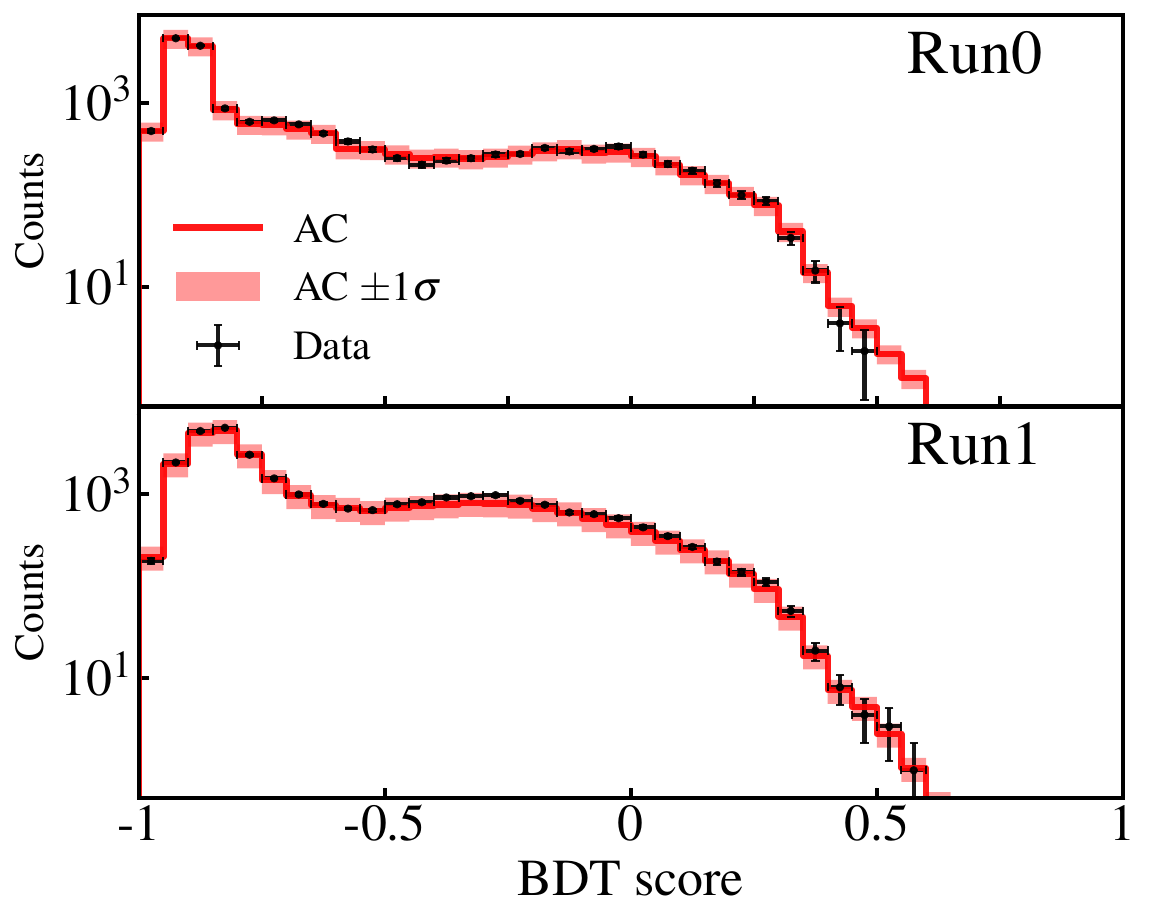}
    \caption{
    The BDT score distributions (top: Run0, bottom: Run1) of the 1-hit sideband data (black data points) overlaid with that from the AC model (red histogram with shaded error band).
    }
    \label{fig:1-hit_sideband}
\end{figure}

\begin{figure}[htp]
    \centering
    \includegraphics[width=0.95\columnwidth]{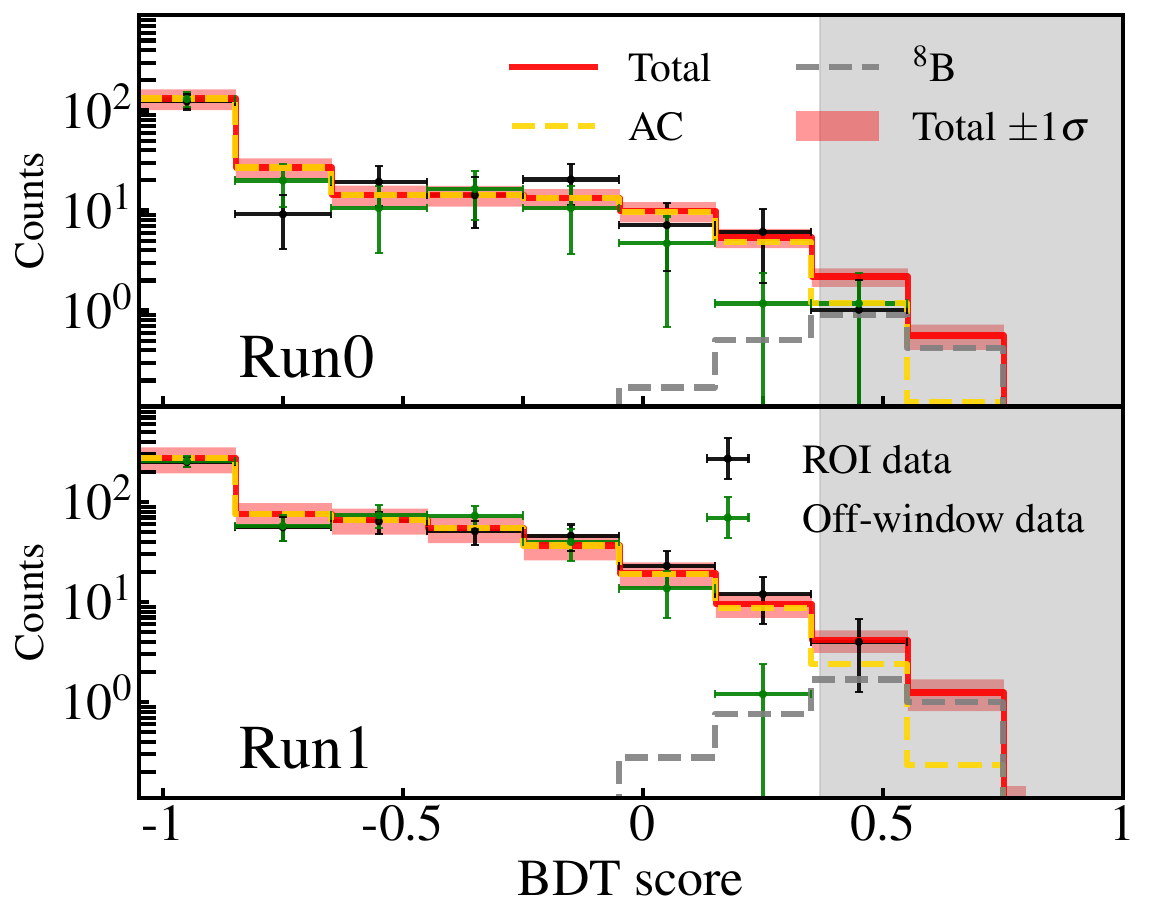}
    \caption{
    The BDT score distributions (top: Run0, bottom: Run1) of the 2- and 3-hit ROI data (black data points) and the off-window data (green data points), overlaid with that from the AC model (gold), B8 signal (gray dashed), and their sum (red), with uncertainty indicated by the shaded bands. The signal region defined by the BDT cut is indicated by the gray-shaded region.
    }
    \label{fig:2-3-hit_roi}
\end{figure}

\textbf{Appendix:}
In this appendix, we provide some details of the rejection of AC background via the BDT method. 
The procedure follows previous analyses on PandaX-II data~\cite{abdukerim2022study} as well as Run0 of PandaX-4T~\cite{ma2023search}.  
The training of the BDT is based on variables built from the $S1$ and $S2$ signals, as well as several noise-related variables based on alternative pulses within 1-$\mu$s window before the $S2$ signal (pre-$S2$ window).
The $S1$ and $S2$ variables include their  charges, waveform shape parameters, and variables related to charge distribution on the PMT arrays. 
Similar variables are defined for alternative pulses in the pre-$S2$ window. 
The scrambled AC background and expected $^8$B signals, both built based on real snippets of data waveforms~\cite{lin2024waveform, abdukerim2022study}, are used to train the BDT. 
Most of the BDT rejection power against the AC background comes from the variables concerning $S2$ signals and pre-$S2$ window pulses. No significant correlation between the $S1$ variables and $S2$ variables (or the pre-$S2$ window variables) is observed. In Fig.~\ref{fig:1-hit_sideband} we compare the BDT score distribution of the real data with $S1$ made up of a single PMT hit (1-hit sideband), expected to be dominated by the AC, against that from our AC model. Excellent agreement is found. Before unblinding, the BDT cut value is optimized to achieve the highest signal sensitivity, under both statistical and background systematic uncertainties. After unblinding, the BDT score distributions of the 2- and 3-hit paired data, and off-window data ($S1$-$S2$ separation exceeding the length of TPC) in Run0 and Run1 are shown in Fig.~\ref{fig:2-3-hit_roi}, together with those from the AC background and B8 signal, normalized based on the best-fit number of events.

\end{document}